\documentstyle[12pt,epsfig]{article}
\newfont{\cmu}{cmu10 scaled\magstep1}
\newcommand{\be}{\begin{equation}}
\newcommand{\ee}{\end{equation}}
\newcommand{\bea}{\begin{eqnarray}}
\newcommand{\eea}{\end{eqnarray}}
\newcommand{\PD}{{\partial}}

\begin{document}
\begin{center}
{\Large\bf Hadron and hadron-cluster production}
\\[.5cm]
{\Large\bf in a hydrodynamical model}
\\[.5cm]
{\Large\bf including particle evaporation}
\\[2cm]
{\bf A.\ Dumitru, C.\ Spieles, H.\ St\"ocker}
\\[0.2cm]
{\small Institut f\"ur Theoretische Physik der J.W.\ Goethe-Universit\"at}\\
{\small Robert-Mayer Str. 10, Postfach 11 19 32}\\
{\small D-60054 Frankfurt a.M., Germany}
\\[0.8cm]
{\bf C. Greiner}
\\[0.2cm]
{\small Institut f\"ur Theoretische Physik der J.\ Liebig-Universit\"at}\\
{\small Heinrich-Buff-Ring 16}\\
{\small D-35392 Giessen, Germany}
\\[1cm]
{\large June 1997}
\end{center}
\begin{abstract}
We discuss the evolution of the mixed phase at RHIC and
SPS within boostinvariant hydrodynamics. In addition to the hydrodynamical
expansion, we also consider evaporation of
particles off the surface of the fluid. The back-reaction of this
evaporation process on the dynamics of the fluid shortens the lifetime of the
mixed phase. In our model this lifetime of the mixed phase is $\le12~fm/c$
in $Au+Au$ at RHIC and $\le6.5~fm/c$ in $Pb+Pb$ at SPS, even in the limit
of vanishing transverse expansion velocity.
Strong separation of strangeness occurs, especially in
events (or at rapidities) with relatively high initial net baryon and
strangeness number, enhancing the multiplicity of MEMOs (multiply strange
nuclear clusters). If antiquarks and antibaryons reach saturation in the
course of the pure QGP or mixed phase, we find that at RHIC the ratio of
antideuterons to deuterons may exceed $0.3$ and even $\overline{^4He}/{^4He}
>0.1$. In $S+Au$ at SPS we find only $\overline{N}/N\approx0.1$.
Due to fluctuations, at RHIC even negative baryon number at
midrapidity is possible in individual events, so that the antibaryon and
antibaryon-cluster yields exceed those of the corresponding baryons and
clusters.
\end{abstract}
\vfill
\begin{flushleft}
{\cmu Work supported by BMBF, DFG, GSI.}
\end{flushleft}
\newpage

\section{Introduction}
The initial (pre-equilibrium) stage of heavy-ion collisions at BNL-RHIC
energies has been intensively studied during the last years within
various models, see e.g.\ refs.\ \cite{Geiger,Eskola,partequ}.
As a consequence of the huge number
of (sea) partons in the inital state and the large cross sections
(especially for $gg$) already after a very short (proper) time $\tau_i
\approx0.1-0.2~fm/c$ the parton momentum space distributions at midrapidity
are isotropic. Their widths
correspond to a very high temperature on the order of $T_i\simeq500~MeV$.
On the other hand, the approach to chemical equilibrium probably takes
somewhat longer. Nevertheless, it appears reasonable to assume that at
midrapidity a quark-gluon plasma (QGP) in thermal and chemical equilibrium
emerges after a few $fm/c$. This QGP subsequently expands and cools. If 
equilibrium is maintained during the expansion and if the
phase transition to the finally observed hadrons is of first order (with
a sufficiently large latent heat), the system will spend a large part
of its lifetime in the mixed phase \cite{soft,RiGy,Barz}.

In this paper we discuss a scenario for the hydrodynamical evolution of
the mixed phase, also including particle evaporation off the surface
of the system and the back-reaction of this
time-dependent freeze-out. We also account for
net strangeness conservation to compute the multiplicities of strange
hadrons and -clusters. General aspects and thermodynamic properties
of strange matter are discussed in \cite{RevStrange,sdist,schaffi2}.

We do not consider the evolution of the pure QGP phase since
most calculations of the
initial conditions indicate that at early times $\tau\le 1-3~fm/c$
the parton gas might not be in chemical
equilibrium, even though it is thermalized \cite{Geiger,Eskola,partequ}.
Also, the lifetime of the mixed phase at RHIC
exceeds that of the pure QGP by roughly a factor of $3$.
Thus, probably many more hadrons are evaporated from the mixed phase.
Anyhow, the emission of hadrons from the pure QGP phase \cite{Kajino}
would further increase the evaporation rates (cf.\ section \ref{TheModel})
and the effects discussed here would be even stronger.

We consider the following dynamical picture. At (proper)
time $\tau_M$ a chemically and thermally equilibrated system
is formed, which is assumed to be in the mixed phase. The mixed phase consists
of a quark-gluon plasma modelled by a MIT bagmodel equation of state
(mass of $s$-quark: $m_s=150~MeV$),
and a hadron gas that includes all well-established (strange and non-strange)
hadrons up to masses of $2~GeV$ (the list is given in \cite{Winck}).
For the hadron gas Hagedorn's eigenvolume correction is applied 
\cite{Hagedorn}. We employ a bag parameter of $B=380~MeV/fm^3$, leading to
a mixed phase temperature (at small $\mu_q$ and $\mu_s$) of
$T_C\approx160~MeV$. The same equation of state was used in \cite{CSCG}
to study hadron ratios, cluster abundancies, and strangeness distillation
in $S+Au$ reactions at CERN-SPS within a similar model 
\cite{sdist,schaffi2,StoeQM95}, see also \cite{Barz}.
However, in contrast
to the model presented here, spherical symmetry was assumed and
boostinvariant hydrodynamical expansion was not taken
into account. Also, the evaporation rates (cf.\ section \ref{TheModel}) are
different in the two models.

Initially the fraction of volume occupied by
quarks and gluons equals one, $\lambda(\tau_M)=V^{QGP}/V^{tot}=1$.
Hydrodynamic expansion and evaporation of hadrons from the surface
set in, leading to a decrease of the total energy-, entropy-, and
net baryon density. Except for the permanent decoupling of hadrons at
the boundary to vacuum, the mixed phase is assumed to remain in
equilibrium while hadronization proceeds.

At some (proper) time $\tau_H$ the quark-gluon
content of the mixed phase (i.e., $\lambda$) vanishes and all energy etc.\
has been converted into hadrons. In principle, the hydrodynamic
evolution continues until the decoupling stage is reached. However,
experimental data for heavy-ion collisions at AGS and SPS indicate
that the chemical potentials and the temperature at the time of decoupling are
close to the phase boundary~\cite{BrM}, i.e.\ that the hydrodynamic
expansion of the hadronic phase can be neglected. We therefore
decouple all remaining hadrons from the hydrodynamic flow immediately after
the conversion of mixed phase matter into hadrons is complete.
Note however, that especially the multiplicity of antibaryons and
anticlusters may depend strongly on the volume and the lifetime of the
hadronic phase (due to annihilation) \cite{AntiB,StoeQM95}, even if it is very
short.

We also employ the same dynamical picture (with different initial conditions,
of course) for $S+Au$ and $Pb+Pb$ collisions at SPS, although the assumptions
of boostinvariant expansion and full strangeness saturation might
only be moderately fulfilled
at this energy. Nevertheless, by comparing our results to those of
ref.\ \cite{CSCG} (evaporation without boostinvariant expansion), we can
give upper and lower limits for the time-scale of chemical freeze-out.

\section{Hydrodynamic expansion including particle evaporation}
\label{TheModel}
The dynamical evolution of the mixed phase is described within a
simple hydrodynamical model which is consistent with the
inside-outside cascade picture at high energies~\cite{Bj}. In
this model, the thermodynamic state of the system is defined along proper time
hyperbolas $\tau = \sqrt{t^2-z^2}$.
The flow-rapidity $\eta$ is assumed to be equal to the space-time
rapidity, i.e.\ the (longitudinal) flow velocity at
the space-time point $(t,z)$ equals $v_\parallel=z/t$.
The space-time volume element is
$d^4x = d^2x_T \tau d\tau d\eta$.
In the case of a perfect fluid (i.e.\ disregarding viscosity and
thermal conductivity), the rapidity density of each quantity
corresponding to a conserved current is (proper) time independent,
\be \label{ConsLaws}
\frac{\PD}{\PD\tau}\frac{\PD A(\tau)}{\PD \eta} = 0\quad.
\ee
In the case without explicit particle evaporation, this equation holds for the
(total) entropy $S$, the net baryon
number $N_B$, and the net strangeness $N_s$. The energy per rapidity
interval decreases due to the work performed by the expansion according to
\be \label{hydroexp}
\frac{\PD}{\PD\tau}\frac{\PD E(\tau)}{\PD \eta} = -p
\frac{\PD}{\PD\tau}\frac{\PD V(\tau)}{\PD \eta}
= -\pi R_T^2 p 
\quad, \label{EnEv}
\ee
where $p$ denotes the pressure. For simplicity, we consider only
longitudinal expansion, i.e.\ $\PD R_T/\PD\tau=0$, assuming that collective
transverse expansion of the mixed phase is slow. The transverse shock
wave, which can additionally produce up to $\approx10\%$ of entropy
(for almost net baryon free matter) \cite{RiGy,shock}, therefore is absent.
A microscopic calculation of entropy production in the phase transition
to hadronic matter can be found in ref.\ \cite{Bertsch}.
In our case, this additional entropy can however be thought of being
absorbed in the initial entropy which has comparable uncertainties
(see next section).

In addition to the hydrodynamic expansion we consider particle evaporation off
the surface of the midrapidity cylinder with transverse radius $R_T$. 
For the evaporation rate in the local restframe of the fluid we estimate
\bea \label{PartEvRate}
\frac{dN^{ev}_i}{d\tau d\eta} &=& O^H \int\frac{d^3p}{E} \,f_i(E)\,
\frac{\vec{p}_T\cdot\vec{r}_T}{r_T}\,\Theta\left(\vec{p}_T\cdot\vec{r}_T
\right)
\\ &=&  O^H n^H_i \frac{1}{\pi}
\left< \frac{p_T}{E}\right>_i \quad, \nonumber
\eea
where $O^H=2\pi R_T\tau$ denotes the surface of the
expanding cylinder.
$n^H_i$ is the density of the hadron species $i$ under consideration.

The evaporated particles carry away entropy, baryon number, strangeness,
and energy:
\bea \label{EvRates}
\frac{dS^{ev}}{d\tau d\eta} &=& \sum_i {\cal S}_i 
\frac{dN^{ev}_i}{d\tau d\eta} \quad, \\
\frac{dN_B^{ev}}{d\tau d\eta} &=& \sum_i {\cal N}_{B,i} 
\frac{dN^{ev}_i}{d\tau d\eta} \quad,\\
\frac{dN_s^{ev}}{d\tau d\eta} &=& \sum_i {\cal N}_{s,i}
\frac{dN^{ev}_i}{d\tau d\eta} \quad. \label{EvNs}
\eea
${\cal S}_i\equiv s_i^H/n_i^H$, ${\cal N}_{s,i}\equiv n_{s,i}^H/n_i^H$, 
and ${\cal N}_{B,i}\equiv n_{B,i}^H/n_i^H$
denote the entropy, strangeness, and baryon number
per particle of the hadron species $i$ (for the given values of
$\mu_s$, $\mu_B$, $T$), respectively.
For the pions and temperatures around $160~MeV$ we compute
$\left< \frac{p_T}{E}\right>_\pi\approx0.75$.
For all other particles, we employ
\be \label{vTparticle}
\left< v_T \right>_i = \sqrt{\frac{T}{m_i}}\quad,
\ee
which is close to the non-relativistic limit
\be\label{vTnrparticle}
\left< v_T \right>_i^{nr} = \sqrt{\frac{\pi T}{2m_i}}\quad.
\ee
By numerical integration of eq.\ (\ref{PartEvRate}) we have found that
in the relevant temperature range (\ref{vTparticle}) gives a better
approximation of the exact value than (\ref{vTnrparticle}), especially for
kaons and nucleons.

The evaporated energy could be computed employing an expression similar
to eqs.\ (\ref{EvRates})-(\ref{EvNs}),
\be \label{ENEvap}
\frac{dE^{ev}}{d\tau d\eta} = \sum_i {\cal E}_i
\frac{dN^{ev}_i}{d\tau d\eta} \quad,
\ee
with ${\cal E}_i\equiv\epsilon_i^H/n_i^H$ denoting the energy per
hadron of species $i$. Eq.\ (\ref{ENEvap}) thus assumes that the energy per
particle in the free-streaming regime is equal to that in the thermal
ensemble. However, this would mean that the evaporation
of hadrons performs mechanical work, i.e., 
\be
-pdV^{ev}= dE^{ev}-TdS^{ev}-\mu_i dN_i^{ev}>0 \quad.
\ee
The volume of the
hadron gas simply shrinks until it disappears, while the volume of the QGP
stays constant. Thus, eqs.\ (\ref{EvRates})-(\ref{ENEvap}) do not lead
to conversion of QGP into hadronic matter, unless transverse expansion is
also taken into account (by adding a term $-2\pi R_T\tau p \PD R_T/\PD\tau$
on the right-hand-side of eq.\ (\ref{hydroexp})). We have found that for
moderate transverse expansion velocities, $\PD R_T/\PD\tau\approx0.2$,
the shrinking of the transverse radius due to evaporation and the
increase due to expansion cancel. The resulting evolution and lifetime of the
mixed phase are very similar to that described below.

We therefore assume
that evaporation causes only heat transfer ($TdS^{ev}<0$) and chemical work
($\mu dN^{ev}<0$) but does not perform mechanical work.
In this case the evaporated energy
is determined by the first law of thermodynamics (with $dV^{ev}=0$),
\be
\frac{dE^{ev}}{d\tau d\eta} = T\frac{dS^{ev}}{d\tau d\eta} +
\mu_s\frac{dN_s^{ev}}{d\tau d\eta} +\mu_q\frac{dN_q^{ev}}{d\tau d\eta}\quad.
\ee

The time evolution of the thermodynamic quantities
due to particle evaporation and hydrodynamic expansion is
\bea \label{rate_eq}
\frac{dA}{d\tau d\eta} &=& -\frac{dA^{ev}}{d\tau d\eta}
\quad\quad\quad (A=S,\, N_s,\, N_B)\quad,\\
\frac{dE}{d\tau d\eta} &=& -\frac{dE^{ev}}{d\tau d\eta}
-\pi R_T^2 p 
\quad.\label{enrate_eq}\eea
These equations generalize (\ref{ConsLaws},\ref{hydroexp}) to account for
the back-reaction of the freeze-out on the dynamics of the fluid.
Note that there are two distinct time scales in this model,
one given by the (longitudinal) hydrodynamic expansion and the other by
the evaporation process.

The preequilibrium parton evolution yields the initial values
of the entropy, the net baryon number, and the net strangeness, i.e.\
$dS(\tau_M)/d\eta$, $dN_B(\tau_M)/d\eta$, and $dN_s(\tau_M)/d\eta$ (see below).
By virtue of the equation of state and the additional requirement that the
system is in the mixed phase with a QGP-fraction of $100\%$ (i.e.,
$\lambda(\tau_M)=1$), the initial energy at midrapidity is also determined.

At later times $\tau>\tau_M$, eqs.\ (\ref{rate_eq},\ref{enrate_eq})
yield $dE/d\eta$, $dS/d\eta$, $dN_B/d\eta$, and $dN_s/d\eta$. From
these four quantities and Gibbs' condition of phase equilibrium,
$p^{QGP}=p^H$, we can compute $V^{tot}$,
$\mu_q$, $\mu_s$, $T$, and the volume fraction of QGP
$\lambda\equiv V^{QGP}/(V^H+V^{QGP})$.

The evaporated particles are assumed to stream freely to the detector. The
measured hadron multiplicities are thus given by the evaporation rates
(\ref{PartEvRate}), integrated over (proper) time. In addition,
at the time $\tau_H$ where the hadronization is complete
($\lambda(\tau_H)=0$), the entire remaining hadronic system is assumed to
decouple instantaneously (i.e., on the hyperbola $\tau=\tau_H$) from thermal
and chemical equilibrium.

\section{Initial conditions}
On average over many events,
the natural choice for the initial net strangeness (at midrapidity)
is $d N_s/d\eta=0$.
In contrast to heavy-ion reactions at fixed-target energies, the net
baryon number at midrapidity is expected to be relatively small at RHIC.
This is due to the large number of secondary partons as compared to the
number of valence quarks. Nevertheless, it is important to account for
nonvanishing $N_B$, especially when studying the enrichment of the
mixed phase with net strangeness. The models RQMD~1.07 \cite{RQMDStopp},
FRITIOF~7.02 \cite{FRITIOF}, and the PCM \cite{Geiger} predict
$dN_B/d\eta=20-35$ in $Au+Au$ at RHIC.

The entropy per baryon created in the early evolution of the parton gas
is $S/N_B=200$ in the PCM \cite{Geiger}.
In ref.\ \cite{Eskola} $S/N_B=110$ was estimated
for the ``final'' (i.e.\ at the time where isentropic expansion sets in)
entropy per net baryon by assuming perturbative QCD interactions (lowest
order) between the partons in the initial state. However, in this
calculation a transverse momentum cut-off (which is necessary to
render the lowest-order pQCD cross sections finite) of $p_0=2~GeV$
was used. As discussed in \cite{Eskola}, this might be appropriate for
LHC but for lower energies like RHIC the saturation of the produced
gluons should be reached already for a smaller cut-off.
Changing $p_0$ from $2~GeV$ to $1~GeV$ increases the minijet cross sections
and thus $S/N_B$ to about $150$ \cite{Karipriv}. Also, the soft
component was assumed to be constant from SPS to RHIC, $\left
(S/N_B\right)_{soft}\approx 50$. Multiplying this number by an ad-hoc
factor of $2$ increases the total $S/N_B$ to about $150$ for
$p_0=2~GeV$ and to $220$ for $p_0=1~GeV$ \cite{Karipriv}.

The RQMD~1.07 model, based on color-string formation and decay, predicts a
central pion multiplicity of $d N_\pi/d\eta=1000$, leading to $S_\pi/N_B
=3.6N_\pi/N_B=170$ \cite{RQMDStopp}.
If we assume that other particles (e.g.\ kaons) contribute about
$10\%-20\%$ to the total entropy, we find a similar number for
$S/N_B$ as in the PCM. The microscopic phase-space model UrQMD \cite{Winck},
which also approaches RHIC from below assuming that color-string
formation and subsequent fragmentation is the dominant mechanism for
secondary production (as at SPS), predicts about
$30$ net baryons, $900$ pions, and $80-100$ kaons at midrapidity in central
($b=2~fm$) $Au+Au$ collisions at RHIC \cite{URQMD}.
Thus, a value of $S/N_B=200$ (at midrapidity) for an
average central $Au+Au$ event at RHIC seems reasonable.

Considering $\tau_M$ (the initial time for the mixed phase), we first note
that once the midrapidity numbers
for $N_B$ and $S$ are fixed, the corresponding coordinate space densities
$n_B$ and $s$ behave like $1/\tau$. Thus, $\tau_M$ must not be chosen
too small for then the system cannot be in the mixed phase initially
(the equation $p^{QGP}=p^H$ cannot be fulfilled). Also, the initial
transverse radius $R_T$, as given
by the thermodynamic relations, depends on
$\tau_M$ (through the volume of the midrapidity region $\PD V/\PD\eta=
\pi R_T^2\tau$). For the parameter set $N_s=0$, $N_B=25$, $S/N_B=200$,
we choose $\tau_M=5~fm/c$ which leads to $R_T=5.4~fm$.
A similar
value for $\tau_M$ would emerge if an equilibrated QGP with temperature
$T\approx500~MeV$ at the initial time $\tau_i=0.1-0.2~fm/c$ was created
\cite{RiGy,Mustafa}.

For collisions at SPS we employ $N_s=0$ initially.
The specific entropy required to fit the particle ratios measured in
$S+Au$ with our equation of state is $S/N_B=45$ \cite{CSCG}. For central
$Pb+Pb$ we employ the same value. In $S+Au$ the net baryon number
at midrapidity, as measured by the NA35 collaboration \cite{NA35NetB}, is
$dN_B/d\eta\approx16$. For $Pb+Pb$ we assume $dN_B/d\eta=80$ which
results in $\approx61$ net nucleons at midrapidity (cf.\ section
\ref{SPSResults}). If we simply scale by $1/2$ (at high energies the
initial isospin asymmetry is ``deposited'' in the pions and the nucleons become
isospin symmetric), the net proton
multiplicity is close to the preliminary NA49 data \cite{NA49}.
For the (proper) time where the evolution of
the mixed phase starts we assume $\tau_M=1.5~fm/c$ ($S+Au$) and
$\tau_M=3~fm/c$ ($Pb+Pb$),
which leads to $R_T=3.7~fm$ and $R_T=5.9~fm$, respectively.

\section{Results for RHIC} \label{RHICResults}
\begin{figure}[htp]
\centerline{\hbox{\epsfig{figure=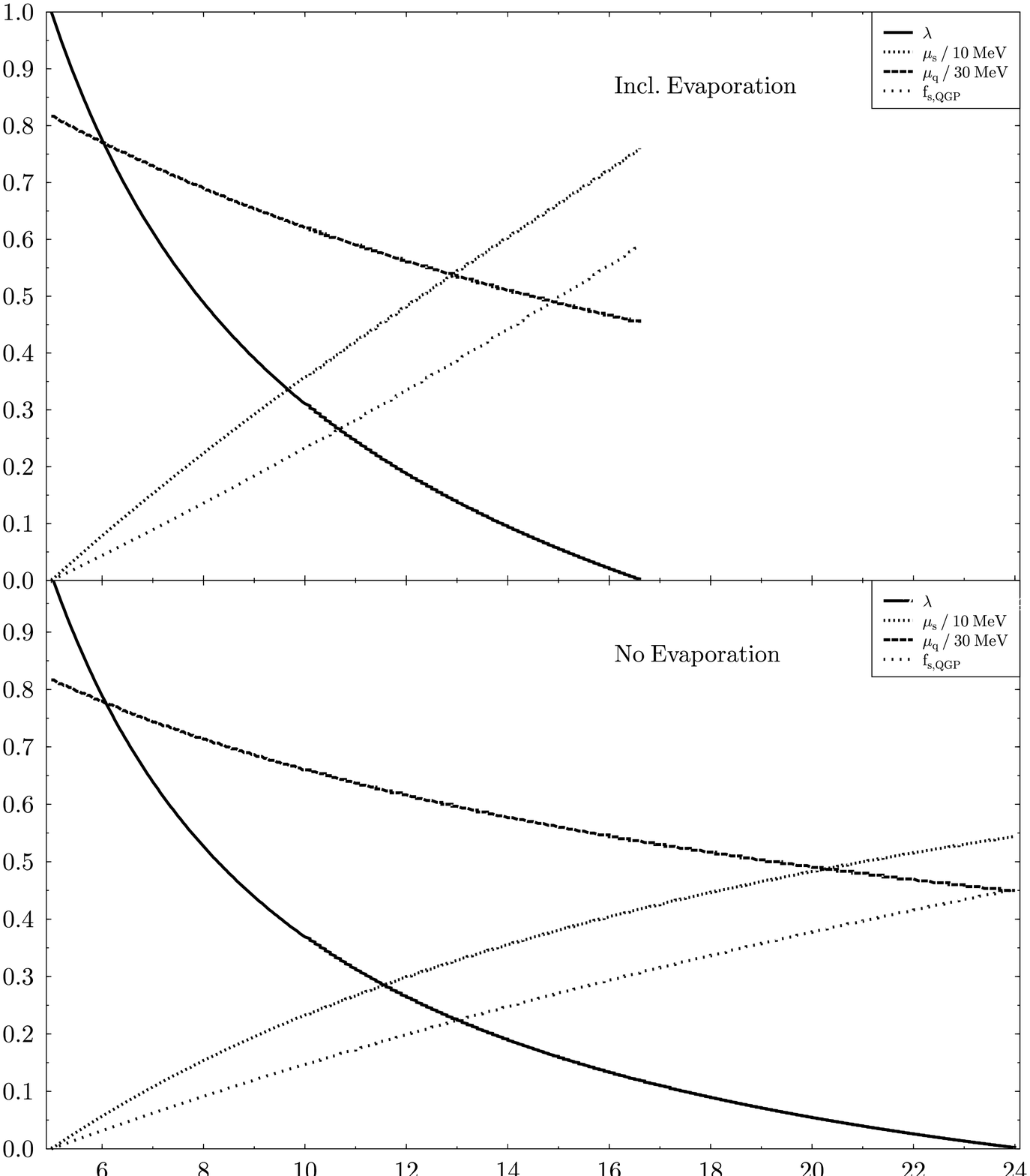,height=15cm,width=12cm}}}
\vspace*{1cm}
\caption{Time evolution of the volume fraction occupied by QGP, the
chemical potentials of $s$- and $u,d$-quarks, and the number of net strange
quarks per baryon in the QGP. The initial
conditions are as in table 1 (average $Au+Au$
at RHIC; bottom:
only longitudinal hydrodynamic expansion without particle evaporation; top:
including evaporation).}
\label{fig1}
\end{figure}

Figure \ref{fig1} shows the time evolution of $\lambda$, $\mu_s$,
$\mu_q$, and $f_{s,QGP}\equiv N_s^{QGP}/N_B^{QGP}$.
Evaporation leads to a faster conversion of QGP into hadronic matter
and reduces the lifetime of the mixed phase from $\tau_H-\tau_M=19~fm/c$
(only longitudinal boostinvariant expansion) to $\tau_H-\tau_M=11.6~fm/c$.
Thus, even if the transverse expansion of the mixed phase is extremely slow 
\cite{RiGy}, the lifetime is considerably shortened by the permanent
decoupling of particles close to the surface.

In contrast to the situation for lower
vacuum pressure $B$ and initial specific entropy
\cite{Barz,sdist,schaffi2,CSCG,Subram},
for $S/N_B\ge40$ and $B=380~MeV/fm^3$ the temperature
in the mixed phase ($T\approx160~MeV$) is essentially constant and does
not increase significantly (although $S^{QGP}/N_B^{QGP}$ does increase).
The variations of the chemical potentials $\mu_s$ and $\mu_q$ in time are
much bigger.
Also, strangeness separation between QGP and hadrons occurs
($f_{s,QGP}>0$) even without evaporation ($N_s=0$ in the total system in this
case).
This is a consequence of the fact that from $N_s=0$ and $N_B>0$ also
$\mu_s>0$ follows if hadrons are present (i.e., $\lambda<1$) \cite{sdist}.
For the initial conditions as in figure \ref{fig1} the evaporation of
strangeness (which is mainly due to kaons) leads to a net strangeness at the
final break-up (i.e.\ at the time where $\lambda=0$) of $N_s/N_B=11.5\%$.

Due to the fact that (for these initial conditions) $K$, $p$, $\Lambda$
are more abundant in the hadron gas than the corresponding antiparticles,
they are evaporated earlier and their average freeze-out times are smaller
than those of the antiparticles \cite{CSCG,GyCorr,soff}. As shown in ref.\
\cite{soff}, for low vacuum pressure $B^{1/4}=160~MeV$
(slow hadronization) this phenomenon affects the properties of the
$K-\overline{K}$ correlation function.
In our model, however, as a consequence of the boostinvariant expansion
most particles are emitted in the breakup of the purely hadronic phase,
washing out this signal completely.
We obtain equal average freeze-out times for $K$ and
$\overline{K}$: $\langle\tau_{fo}\rangle_{K,\overline{K}}=15~fm/c$
(these numbers include the kaons from the final breakup but not those from
decays of higher-mass resonances).

\begin{table}{}
\begin{center}
\begin{tabular}{|ccccccc|} \hline
$\stackrel{ }{\pi}$ & $\stackrel{ }{K}$ & $\stackrel{ }{\overline{K}}$ & 
$\stackrel{ }{N}$ & $\stackrel{ }{\overline{N}}$ 
& $\stackrel{ }{\Lambda}$ & $\stackrel{ }{\overline{\Lambda}}$\\ \hline
830 & 100 & 92.6 & 44.4 & 25.8 & 13.2  & 8.67\\
267 &  31 & 27.1 & 11.4 &  5.97&  3.19 & 2.05\\
\hline
$\stackrel{ }{d}$ & $\stackrel{ }{\overline{d}}$ & $\stackrel{ }{^4He}$ & 
$\stackrel{ }{\overline{^4He}}$ & $\stackrel{ }{\Lambda\Lambda}$ & & \\ 
\hline
0.104 & 0.0355 & $1.20\cdot10^{-6}$ & $1.43\cdot10^{-7}$ & 0.0051 & & \\
0.024 & 0.0066 & $2.38\cdot10^{-7}$ & $1.79\cdot10^{-8}$ & 0.0010 & & \\
\hline
\end{tabular}\\[1ex]
\end{center}
{Table 1: Multiplicities of hadrons and clusters emitted from the
expanding midrapidity cylinder.
Feeding from decays of higher-mass resonances is included.
The first line shows the total multiplicity (evaporated hadrons plus
those from the final breakup at time $\tau=\tau_H$) while the second
line shows only the evaporated particles.
Initial conditions: $dS/d\eta\,/\,dN_B/d\eta=200$,
$dN_B/d\eta=25$, $dN_s/d\eta=0$, $\tau_M=5~fm/c$
($Au+Au$ at RHIC).}
\label{tab1}
\end{table}

\begin{table}{}
\begin{center}
\begin{tabular}{|ccccccc|} \hline
$\stackrel{ }{\pi}$ & $\stackrel{ }{K}$ & $\stackrel{ }{\overline{K}}$ & 
$\stackrel{ }{N}$ & $\stackrel{ }{\overline{N}}$ & 
$\stackrel{ }{\Lambda}$ & $\stackrel{ }{\overline{\Lambda}}$\\ \hline
821 &  82.6 & 108 & 51.5 & 20.9 & 19.7  & 5.49\\
265 &  25.7 & 31.7& 13.2 & 4.92 &  4.68 & 1.32\\
\hline
$\stackrel{ }{d}$ & $\stackrel{ }{\overline{d}}$ & $\stackrel{ }{^4He}$ & 
$\stackrel{ }{\overline{^4He}}$ & $\stackrel{ }{\Lambda\Lambda}$ & & \\ 
\hline
0.144 & 0.0240 & $2.30\cdot10^{-6}$ & $6.54\cdot10^{-8}$ & 0.0102 & & \\
0.033 & 0.0045 & $4.36\cdot10^{-7}$ & $8.47\cdot10^{-9}$ & 0.0020 & & \\
\hline
\end{tabular}\\[1ex]
\end{center}
{Table 2: Same as in table 1 but for the initial condition
$dS/d\eta\,/\,dN_B/d\eta=100$, $dN_s/d\eta=dN_B/d\eta=50$, $\tau_M=5~fm/c$
($Au+Au$ at RHIC).}
\label{tab2}
\end{table}
The initial conditions in individual events may strongly deviate
from our standard set. There may well be large
fluctuations \cite{Spieles} in $N_s$ as well as in the net baryon number
$N_B$, and the entropy per net baryon, $S/N_B$, if one considers
only a small subinterval (e.g.\ $\pm0.5$ around midrapidity) of the total
rapidity gap. We assume that even in such events boostinvariance around
midrapidity is approximately valid.

In the tables we compile
particle and cluster abundancies for our standard initial conditions
and a calculation with higher initial net baryon number and an excess
of $s$-quarks. The first line gives the sum over evaporation
and final breakup.
Below, the contribution from evaporation only
(without final breakup) is quoted. This latter part of antibaryons and
clusters survives even if the space-time volume of the
chemically equilibrated purely hadronic phase would be significant (which
would lead to cooling below $T_C$).
However, as already mentioned in the introduction, this is probably not the
case \cite{CSCG,BrM,Bertsch}. The pion multiplicity is perhaps somewhat
underpredicted, which may be due to the fact that the effective volume
correction should be done different for pions \cite{Gorenstein}.

For the standard set of initial conditions (table 1),
the higher initial specific entropy
(as compared to BNL-AGS and CERN-SPS) leads to considerable antibaryon
production ($\overline{N}/N$, $\overline{\Lambda}/\Lambda>0.5$).
In $S+Au$ at SPS one finds only
$\overline{N}/N\approx0.1$. At RHIC, even the ratio of deuterons to
antideuterons and helium to antihelium is larger than $0.1$.

For the second set of initial conditions (table 2), of course, the initial
values of $\mu_s$ and $\mu_q$ are considerably higher than in fig.\
\ref{fig1}. The higher $\mu_q$-values enhance the multiplicity
of deuterons by $40\%$ and that of $^4He$ by a factor of $2$.
Also, a strong strangeness separation between the QGP and the
hadrons in the mixed phase occurs, leading to $f_{s,QGP}$-values up to
$\approx1.44$. For example, in this calculation $\mu_s$ increases from
$37~MeV$ initially to $48~MeV$ at the time where the transition to purely
hadronic matter is complete. These $\mu_s$-values are comparable to those
extracted in ref.\ \cite{CSCG} for $S+Au$ collisions at SPS (cf.\ fig.\
\ref{fig2}), while at the same time $\mu_q$ is considerably lower.
This can be clearly observed in the
ratio $\Lambda/\overline{\Lambda}=3.6$, which is larger than
$N/\overline{N}=2.5$ ($S+Au$ at SPS: $N/\overline{N}\approx9$,
while $\Lambda/\overline{\Lambda}\approx5$).

If experiments were able to trigger on events
(or rapidities) with high net baryon and strangeness density, the
detection of the so-called MEMOs (meta stable
multiple strange objects) \cite{schaffi2,MEMO} might be possible.
As an example, note that for our second set of initial conditions the
multiplicity of the lightest MEMO, the $\Lambda\Lambda$-cluster
(for which we assume a mass of $m_{\Lambda\Lambda}=2200~MeV$, i.e.\
a binding energy of $30~MeV$), increases by a factor of $2$. A similar
amount of $\Lambda\Lambda$-clusters as in $Pb+Pb$ at SPS is emitted (cf.\
table 3), however, from a larger volume.

Since at RHIC the initial net baryon number is not very far from zero, even
events with an excess of antiquarks ($N_B<0$) may not be too rare.
The initial conditions $S=5000$, $N_s=0$, $N_B=-25$, $\tau_M=5~fm$
lead to the same particle and cluster yields as in table 1, except that
particles and antiparticles have to be interchanged. As a result, the
multiplicities of antibaryon clusters are enhanced by factors of $\approx3$
for antideuterons and $\approx8$ for antihelium.

\section{Results for SPS} \label{SPSResults}
\begin{figure}[htp]
\centerline{\hbox{\epsfig{figure=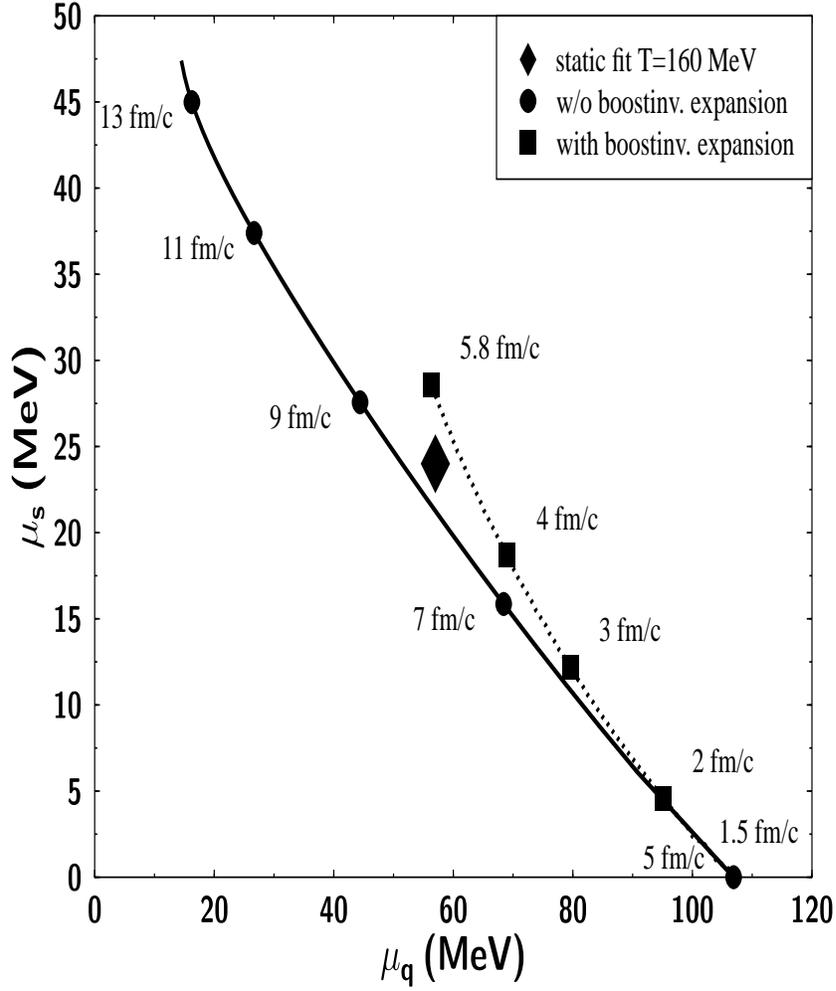,height=15cm,width=12cm}}}
\vspace*{1cm}
\caption{Time evolution of $\mu_s$ and $\mu_q$ in the evaporation model
of ref.\ \cite{CSCG} (without boostinvariant hydrodynamic expansion),
compared to our model. Our initial conditions for the midrapidity cylinder
are $dS/d\eta\,/\,dN_B/d\eta=45$, $dN_B/d\eta=16$, $dN_s/d\eta=0$,
$\tau_M=1.5~fm/c$ ($S+Au$ at SPS).
The parameters of the static fit are also indicated \cite{CSCG}.}
\label{fig2}
\end{figure}

As shown in ref.\ \cite{CSCG}, the hadron ratios measured in
$S+Au$-collisions at SPS can be fitted with our equation of state
assuming isochronous chemical freeze-out (as in \cite{BrM}) at
$T=160~MeV$, $\mu_S=24~MeV$, and $\mu_q=56~MeV$ (indicated by the diamond
in fig.\ \ref{fig2}). On the other hand, a reasonable fit was also
obtained assuming formation of a mixed phase in equilibrium, which then
hadronizes by evaporating hadrons from the surface,
until the whole entropy, baryon number etc.\ is converted into free-streaming
hadrons. Due to the neglect
of hydrodynamic expansion, in this latter model the lifetime of the
mixed phase, and thus the time interval in which the hadrons freeze out
\cite{CSCG,soff}, is relatively large ($8-9~fm/c$, cf.\ fig.\
\ref{fig2}). In contrast to the isochronous freeze-out scenario, the
average freeze-out times for the various hadron species differ considerably.
This is due to the fact that $\mu_s$ and $\mu_q$ are strongly time
dependent and a transition to purely hadronic matter (which freezes out
more or less instantaneously) does not take place.

Here, in our present model (including boostinvariant longitudinal expansion)
the lifetime of the mixed phase is only $4.3~fm/c$, the midrapidity
region then enters the purely hadronic phase and breaks up. As can be seen
in fig.\ \ref{fig2}, this final breakup occurs close to the `static-fit'
point. Most hadrons are emitted at this point, only few
are evaporated\footnote{Therefore,
it is obvious that the (midrapidity) particle ratios in our model are
very close to those quoted in refs.\ \cite{CSCG,BrM} and we do not need to
repeat this discussion here.}. In the other
extreme case (only longitudinal expansion without evaporation) the
lifetime of the mixed phase for our equation of state is $\tau_H-
\tau_M=5.7~fm/c$ and the breakup at the time $\tau_H=7.2~fm/c$ occurs
exactly at the `static-fit' point in fig.\ \ref{fig2}.
Our model thus is close to the sudden
freeze-out scenario (at some specific (proper) time) and the freeze-out times
of the various hadron species do not differ significantly.
We repeat that for the considered high specific entropies and bag pressures
the temperature is essentially time independent, $T=158\pm2~MeV$, while
the chemical potentials $\mu_s$ and $\mu_q$ vary considerably.
\begin{table}{}
\begin{center}
\begin{tabular}{|ccccccc|} \hline
$\stackrel{ }{\pi}$ & $\stackrel{ }{K}$ & $\stackrel{ }{\overline{K}}$ & 
$\stackrel{ }{N}$ & $\stackrel{ }{\overline{N}}$ & 
$\stackrel{ }{\Lambda}$ & $\stackrel{ }{\overline{\Lambda}}$\\ \hline
570 &  75.9 &  53.4&  68.1&   7.48&  16.7&   3.04\\
130 &  18.1 &  10.0&  14.3&    1.00&   2.88&    0.463\\
\hline
$\stackrel{ }{d}$ & $\stackrel{ }{\overline{d}}$ & $\stackrel{ }{^4He}$ & 
$\stackrel{ }{\overline{^4He}}$ & $\stackrel{ }{\Lambda\Lambda}$ & & \\ 
\hline
0.373 & $4.57\cdot10^{-3}$ & $2.44\cdot10^{-5}$ & $3.54\cdot10^{-9}$ &
0.0130 & &\\
0.0873 & $4.20\cdot10^{-4}$& $8.42\cdot10^{-6}$& $1.71\cdot10^{-10}$ &
$2.04\cdot10^{-3}$ & & \\
\hline
\end{tabular}\\[1ex]
\end{center}
{Table 3: Same as in table 1 but for the initial condition
$dS/d\eta\,/\,dN_B/d\eta=45$, $dN_B/d\eta=80$, $dN_s/d\eta=0$,
$\tau_M=3~fm/c$ ($Pb+Pb$ at SPS).}
\label{tab3}
\end{table}

\begin{figure}[htp]
\centerline{\hbox{\epsfig{figure=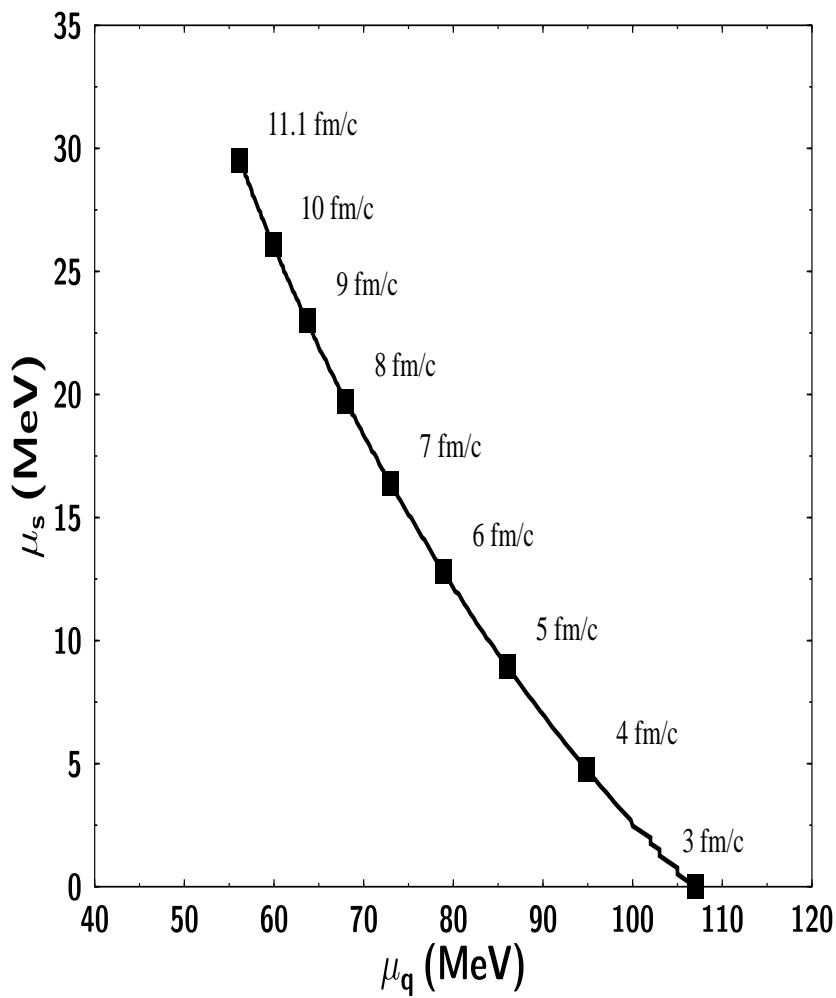,height=15cm,width=12cm}}}
\vspace*{1cm}
\caption{Time evolution of $\mu_s$ and $\mu_q$
in our model (including evaporation and boostinvariant expansion). 
Initial conditions for the midrapidity cylinder as in table 3
($Pb+Pb$ at SPS).}
\label{fig3}
\end{figure}

In $Pb+Pb$ collisions, the lifetime of the mixed phase almost doubles
(fig.\ \ref{fig3}) but still is $2-3~fm/c$ below
the one at RHIC. The resulting hadron multiplicities are shown in table 3.
The pion and net nucleon multiplicities fit well to the preliminary
NA49 data \cite{NA49}. Assuming isospin symmetry for the nucleons,
$\overline{p}=\overline{N}/2$, we find
a ratio $\overline{p}/\overline{\Lambda}=2.5$.
Due to the higher $\mu_q$ values, $d$ and $^4He$ are
more abundant than at RHIC, while the corresponding antinuclei are strongly
suppressed. Note that we obtain few less deuterons but many more
helium nuclei than in the UrQMD model \cite{URQMD}.
One also observes that the fraction of hadrons that are
evaporated directly from the mixed phase (i.e. without those emitted at
the breakup of the purely hadronic phase) is higher at RHIC than
at SPS. The longer the lifetime of the mixed phase is, the more important
evaporation gets. At LHC, a large part of (the entropy of) the mixed phase
will probably be evaporated (i.e.\ directly converted into free-streaming
hadrons) before the fluid reaches the purely hadronic phase.

\section{Summary}
In this paper we discussed the evolution of the mixed phase presumably
created in heavy-ion collisions at SPS and RHIC,
employing a model that combines (boostinvariant) longitudinal hydrodynamical
expansion and particle evaporation off the surface of the fluid.
For reasonable choices of the initial conditions, our model reproduces
the hadron ratios measured in $S+Au$ at SPS (as compiled in \cite{BrM}).
The interplay of the evaporation process and the hydrodynamical
expansion (and vice versa) leads to considerably shorter lifetimes of the
mixed phase (i.e.\ to faster hadronization) as compared to the scenarios
without evaporation (only hydrodynamical expansion
\cite{RiGy,Bj,Mustafa,Schlei}) or
without hydrodynamical expansion (only evaporation \cite{sdist,CSCG}).

We have also calculated the multiplicities of various hadrons, clusters,
and hyperon-clusters in $Pb+Pb$ at SPS and $Au+Au$ at RHIC (our results
for $S+Au$ at SPS are not presented here since they are similar to those of
ref.\ \cite{CSCG}). At SPS, $\mu_q$ is large (and in particular larger
than $\mu_s$), implying that nucleon to antinucleon and hyperon to
antihyperon ratios are big ($\ge5$), in contrast to RHIC, where antinucleons
and antihyperons are more abundant than at SPS by a factor of $3$.

At RHIC, the initial conditions of the central region may fluctuate.
This offers the very interesting opportunity to produce clusters of
antibaryons and hyperons. In events with unusually high
net baryon density or net strangeness (at midrapidity) the multiplicity
of $d$, $^4He$ and MEMOs may be enhanced by up to $100\%$ (as compared
to the average). In this case $\mu_s>\mu_q$ can be reached, leading to
an emission of more hyperons per antihyperon than of nucleons per antinucleon.
On the other hand, negative net baryon number (excess of antiquarks) at
midrapidity increases antinuclei abundancies by up to a factor
of $10$.
\\[1cm]
{\bf Acknowledgements:} We thank D.\ Ardouin and D.H.\ Rischke for helpful
discussions, and K.J. Eskola for
clearifying comments on entropy production in the preequilibrium stage.

\end{document}